\documentclass[12pt,reqno]{amsart}

\usepackage{a4}
\usepackage{amsmath, amssymb}              
\usepackage{graphicx}                      

\theoremstyle{plain}            
\newtheorem{theorem}{Theorem}[section]
\newtheorem*{theorem*}{Theorem}

\newtheorem{lemma}[theorem]{Lemma}

\theoremstyle{definition}       

\theoremstyle{remark}           
\newtheorem{remark}[theorem]{Remark}
\newtheorem{example}[theorem]{Example}

\numberwithin{equation}{section}

\DeclareMathOperator{\dist}   {dist}
\DeclareMathOperator{\dom}    {dom}

\DeclareMathOperator{\Hom}    {Hom}
\DeclareMathOperator{\spec}   {spec}
\DeclareMathOperator{\supp}   {supp}
\DeclareMathOperator{\vol}    {vol}

\newcommand{\R}{\mathbb{R}}                 
\newcommand{\C}{\mathbb{C}}                 
\newcommand{\N}{\mathbb{N}}                 
\newcommand{\Z}{\mathbb{Z}}                 
\newcommand{\Sphere}{\mathbb{S}}            
\newcommand{\Torus}{\mathbb{T}}             
\newcommand{\eps}{\varepsilon}              
\renewcommand{\phi}{\varphi}                
\newcommand{\e}{\mathrm e}                  
\newcommand{\im}{\mathrm i}                 
\newcommand{\dd}{\mathrm d}                 

\newcommand{\HS}{\mathcal H}                

\newcommand{\per}[1]{\mathcal {#1}}         
\newcommand{\Mper}{\per M}                  
\newcommand{\Nper}{\per N}                  
\newcommand{\Xper}{\per X}                  
\newcommand{\Meps}{{M_\eps}}                
\newcommand{\Mepsper}{{\Mper_\eps}}         
\newcommand{\Neps}{{N_\eps}}                
\newcommand{\Nepsper}{{\Nper_\eps}}         
\newcommand{\Aeps}{{A_\eps}}                
\newcommand{\Ceps}{{C_\eps}}                
\newcommand{\Xeps}{{X_\eps}}                

\newcommand{\Neu}{{\mathrm N}}              
\newcommand{\Dir}{{\mathrm D}}              
\newcommand{\laplacian}[1]{\Delta_{{#1}}}   
\newcommand{\laplacianD}[1]{\Delta^\Dir_{{#1}}}
\newcommand{\laplacianN}[1]{\Delta^\Neu_{{#1}}}
\newcommand{\laplacianT}[1]{\Delta^{\theta}_{{#1}}}
\newcommand{\EW}[2]{\lambda_{#1}({#2})}     
\newcommand{\EWD}[2]{\lambda^\Dir_{#1}({#2})}
\newcommand{\EWN}[2]{\lambda^\Neu_{#1}({#2})}
\newcommand{\EWT}[2]{\lambda^{\theta}_{#1}({#2})}
\newcommand{\EWND}[2]{\lambda^{\Neu,\Dir}_{#1}({#2})}

\newcommand{\Ci} [1]{C^\infty ({#1})}       
\newcommand{\Cci}[1]{C_{\mathrm c}^\infty ({#1})} 
\newcommand{\Lsqr}[1]{L_2({#1})}            
\newcommand{\Sob}[2][1]{\HS^{#1}({#2})}     
\newcommand{\norm}[2][{}]{\|{#2}\|_{#1}}    
\newcommand{\normsqr}[2][{}]{\|{#2}\|^2_{#1}} 
\newcommand{\iprod}[3][{}]{\langle{#2},{#3}\rangle_{#1}}    
\newcommand{\bd}  {\partial}                
\newcommand{\dcup}{\dot \cup}               
\newcommand{\restr}[1]{{\restriction}_{#1}} 
\newcommand{\normder}{\partial_\mathrm{n}}  

\begin{document}

\title{Periodic Manifolds with Spectral Gaps}

\author{Olaf Post}

\address{Institut f\"ur Reine und Angewandte Mathematik,
       Rheinisch-Westf\"alische Technische Hochschule Aachen,
       Templergraben 55,
       52062 Aachen,
       Germany}
\email{post@iram.rwth-aachen.de}

\date{05.12.2001}


\begin{abstract}
  We investigate spectral properties of the Laplace operator on a
  class of non-compact Riemannian manifolds. For a given number $N$ we
  construct periodic manifolds such that the essential spectrum of the
  corresponding Laplacian has at least $N$ open gaps.  We use two different
  methods.  First, we construct a periodic manifold starting from an infinite
  number of copies of a compact manifold, connected by small cylinders. In the
  second construction we begin with a periodic manifold which will be
  conformally deformed. In both constructions, a decoupling of the different
  period cells is responsible for the gaps.
\end{abstract}



\maketitle


\section{Introduction}

The spectra of periodic Schr\"odinger or divergence type operators have been
extensively studied. In particular, it is well-known that the spectrum of a
periodic elliptic operator on $L_2(\R^d)$ with smooth coefficients is the
locally finite union of compact intervals, called \emph{(spectral) bands}. We
are mainly interested in the question whether these bands are separated by
\emph{spectral gaps} or not.  By a \emph{gap} in the (essential) spectrum of a
positive operator $H$ we mean an interval $(a,b)$ such that
\begin{displaymath}
  (a,b) \cap \spec H = \emptyset.
\end{displaymath}
To exclude trivial cases we assume that $a$ is greater than the infimum of the
essential spectrum of $H$. The number of gaps is given by the number of
components of the intersection of the resolvent set $\C \setminus \spec H$
with $\R$. Results on spectral gaps where $H$ is a Schr\"odinger or divergence
type operator can be found for example in~\cite{hempel:92b, hempel-herbst:95a,
  hempel-herbst:95b, hempel-lienau:00} (see also the references therein).

In this paper, we want to give examples of (non-compact) periodic manifolds
$\Mper$ such that the corresponding Laplacian $\laplacian \Mper$
\emph{without} potential has spectral gaps. Here, \emph{periodicity} means
that a finitely generated abelian group $\Gamma$ acts isometrically and
properly discontinuously on $\Mper$ (cf.\ for example~\cite{atiyah:76,
  bruening:92, chavel:93, donnelly:81, sunada:89}, periodic manifold are also
called \emph{covering manifolds}). Therefore we obtain the same qualitative
results \emph{only} by the periodic geometry.

As in the Schr\"odinger operator case a decoupling of the different period
cells is responsible for the gaps.  In the Schr\"odinger operator case, the
decoupling is achieved by a high potential barrier separating each period cell
from the others. In the geometric case, decoupling means that the junction
between two period cells is small. Here, a period cell is the closure of a
fundamental domain (see the next section).

From a physical point of view the Laplacian on a manifold is the Hamiltonian
of an electron confined to this manifold (at least in a semi-classical limit,
cf.~for example~\cite{froese-herbst:00, mitchell:01}). Periodic structures
like a periodically curved cylinder or a quantum wire could have applications
in solid state physics. A quantum wire (or quantum wave guide) is a planar
strip, cf.~for example~\cite{exner-seba:89}. The knowledge of the
band-gap structure of $\spec H$ is important for the conductive properties of
the periodic material described by the Hamiltonian $H$. In particular, the
existence and size of the first gap decide whether the material is a conductor
or an insulator.

\subsection*{Basic ideas and results}
The construction of $\Gamma$-periodic manifolds with spectral gaps will be
given later on in detail (cf.~Section~\ref{sec:constr}). Here, we sketch the
\begin{figure}[ht]
  \begin{center}
    \leavevmode
    \begin{picture}(0,0)
       \includegraphics{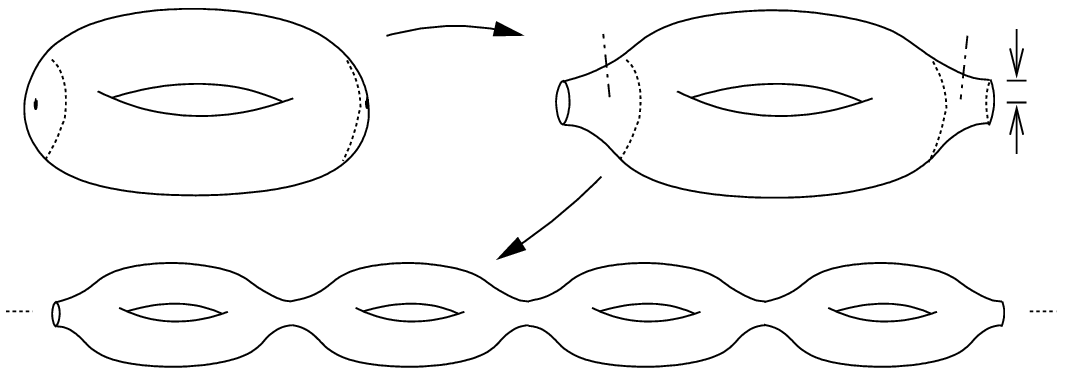}
    \end{picture}
    \setlength{\unitlength}{4144sp}%
    \begin{picture}(4839,1797)(34,-1006)
      \put(421, -1){$\Xeps$}
      \put(1636,-171){$X$}
      \put(3066,  4){ $\Xeps$}
      \put(4366,-106){$\Meps$}
      \put(4279,659){$A_\eps^2$}
      \put(2656,659){$A_\eps^1$}
      \put(4501,-961){$\Mepsper$}
      \put(4816,254){$\eps$}
    \end{picture}
    \caption{Construction of the period cell $\Meps$ and the periodic manifold
     $\Mepsper$ with $\Gamma=\Z$.}
    \label{fig:gaps1}
  \end{center}
\end{figure}
ideas and fix the notation. We start our construction from a compact
Riemannian manifold $X$ of dimension $d \ge 2$ (for simplicity without
boundary). Here, $\Gamma$ is an abelian group with $r$ generators. We choose
$2r$ different points $x_1,\dots,x_{2r} \in X$ and endow each point $x_i$ with
a cylindrical end $A_\eps^i$ (the boundary being isometric to a sphere of
radius $\eps>0$). We call the resulting manifold $\Meps$.
By glueing together $\Gamma$ copies of the period cell $\Meps$ we obtain a
$\Gamma$-periodic manifold $\Mepsper$ (see Figure~\ref{fig:gaps1} for $r=1$
generators). Our first result is the following:
\begin{theorem}
  \label{thm:gaps.constr}
  Each spectral band of the periodic Laplacian $\laplacian \Mepsper$ on
  $\Mepsper$ tends to an eigenvalue of the Laplacian $\laplacian X$ on $X$ as
  $\eps \to 0$. In particular, for each $N \in \N$ there exist at least $N$
  gaps in the spectrum of $\laplacian \Mepsper$ provided $\eps$ is small
  enough.
\end{theorem}
We also prove a similar result in the case where the cells $\Meps$ are joined
by long thin cylinders of fixed length. Let $\Neps$ be the manifold obtained
from $\Meps$ by identifying $\bd A_\eps^{2i-1}$ with a component of the
boundary of a cylinder $C_\eps^i$ of length $L_i \ge 0$ and radius $\eps>0$.
The periodic manifold $\Nepsper$ is obtained in the same way by glueing
together $\Gamma$ copies of $\Neps$.
\begin{theorem}
  \label{thm:gaps.cyl}
  Each spectral band of the periodic Laplacian $\laplacian \Nepsper$ on
  $\Nepsper$ tends to an eigenvalue of the Laplacian $\laplacian X$ on $X$ or
  to an eigenvalue of the Laplacian with Dirichlet boundary conditions on
  $[0,L_i]$ for an $i=1,\dots,r$ with $L_i>0$ as $\eps \to 0$. In particular,
  for each $N \in \N$ there exist at least $N$ gaps in the spectrum of
  $\laplacian \Nepsper$ provided $\eps$ is small enough.
\end{theorem}
The proofs of these two results will be given in Section~\ref{sec:constr}.
Both results are related to articles of Chavel and
Feldman~\cite{chavel-feldman:81} and Ann\'e~\cite{anne:87} where compact
manifolds with small handles resp.\ compact manifolds joined by small
cylinders are analysed. Note that the joined manifolds in~\cite{anne:87} are
not smooth in contrast to our construction.

The second construction is in some sense the reverse of the first construction.  Starting with a given periodic
manifold $\Mper$ of dimension $d \ge 2$ we deform the (periodic) metric $g$ by
a (periodic) conformal factor $\rho_\eps$ to obtain spectral gaps in the
spectrum of the Laplacian. The idea is to let the conformal factor converge to
the indicator function of a closed periodic set $\Xper = \bigcup_{\gamma \in
  \Gamma} \gamma X$ where $X \subset \Mper$ is a closed subset disjoint from
all translates $\gamma X$, $\gamma \ne 0$. This convergence is of course not
uniform because of the discontinuity of the indicator function. We denote the
manifold $\Mper$ with metric $\rho_\eps^2 g$ by $\Mepsper$. We have the
following result:
\begin{theorem}
  \label{thm:gaps.conf}
  Suppose that the dimension of $\Mper$ is greater or equal to $3$.  Then each
  band of the periodic Laplacian $\laplacian \Mepsper$ on the conformally
  deformed periodic manifold $\Mepsper$ tends to an eigenvalue of the Neumann
  Laplacian $\laplacianN X$ on $X$ as $\eps \to 0$. In particular, for each $N
  \in \N$ there exist at least $N$ gaps in the spectrum of $\laplacian
  \Mepsper$ provided $\eps$ is small enough.
\end{theorem}
The assumption $d \ge 3$ is essential here, since in dimension $2$ it is no
longer true that the first band of $\laplacian \Mepsper$ tends to a single
point as $\eps \to 0$. Even in the limit case $\eps=0$, a nontrivial interval
remains due to the special structure of the conformal Laplacian, cf.\ 
Equation~\eqref{eq:quot.dim2}.  Nevertheless, we can prove the existence of
gaps in a simple example by direct calculations (see Example~\ref{ex:dim.2}).
The proof of Theorem~\ref{thm:gaps.conf} and the example in dimension $2$ can
be found in Section~\ref{sec:conf}.

The proofs of our theorems basically use the variational max-min
characterisation of the eigenvalues of an operator with purely discrete
spectrum (called Min-max Principle) to compare the eigenvalues of operators
defined on parameter-depending Hilbert spaces (see the Main
Lemma~\ref{lem:main.lemma}).  The basic idea is taken from \cite{anne:87} and
\cite{fukaya:87} even if the Main Lemma is more general.  Furthermore, Floquet
Theory allows to analyse the spectrum of a periodic operator. Details are
given in the next section.

Davies and Harrell~II~\cite{davies-harrell:87} proved the existence of at
least one gap in the periodic conformally flat case (transforming the
conformal Laplacian in a corresponding Schr\"odinger operator). This result is
a special case of Theorem~\ref{thm:gaps.conf}. Using similar methods, Green
showed in~\cite{green:97} the existence of a finite number of gaps in the
$2$-dimensional conformally flat case. Furthermore,
Yoshitomi~\cite{yoshitomi:98} proved the existence of spectral gaps for the
(Dirichlet) Laplacian on periodically curved quantum wave guides. In all these
three papers the existence of gaps is established basically by analysing a
one-dimensional problem. Here, in contrast, we directly study the
multi-dimensional problem.

Green conjectured that a necessary requirement for a large number of gaps is
for the curvature to be large in absolute value at some points. In the
examples given here the same phenomenon occurs (see
Remarks~\ref{rem:constr.curv} and~\ref{rem:conf.curv}).  Therefore the results
of Fukaya~\cite{fukaya:87} cannot be applied here. Fukaya showed the
continuity of the $k$-th eigenvalue of the Laplacian on $\Meps$ where $\Meps$
is a convergent family of manifolds (in a certain sense) with given bound on
the curvature.

There are also results on periodic operators on manifolds with
\emph{non-commutative} groups $\Gamma$. For example, Br\"uning and Sunada
proved in~\cite{bruening:92} that the Laplacian on a periodic manifold still
has band structure even for certain non-commutative groups $\Gamma$.
Furthermore, Sunada~\cite{sunada:89} showed that --- in contrast to the
Schr\"odinger operator case on $\R^d$ (cf.~\cite{reed-simon-4}) --- there
exist (non-compact!) periodic manifolds with commutative group $\Gamma$ such
that the corresponding Laplacian has an eigenvalue (possibly embedded in
another band).

Finally note that Lott~\cite{lott:01} has constructed a (non-periodic)
$2$-di\-mensio\-nal complete non-compact finite-volume manifold such that the
corresponding Laplacian has an infinite number of gaps.  In the periodic case
in contrast we would expect that the Generalized Bethe-Sommerfeld conjecture
is true, i.e. that there are only finitely many gaps in the spectrum of
$\laplacian \Mper$ if $d \ge 2$.  Skriganov~\cite{skriganov:79} proved this
conjecture for periodic Schr\"odinger operators in Euclidean space.
Furthermore, an asymptotic upper bound on the number of gaps have been
established in~\cite{bruening:92} (not implying the Generalized
Bethe-Sommerfeld conjecture). Note that our results do not say anything
whether the number of gaps is finite or not.

\section{Preliminaries}

\subsection*{Laplacian on a manifold}

Throughout this article we study manifolds of dimension $d \ge 2$. For a
Riemannian manifold $M$ (compact or not) without boundary we denote by $\Lsqr
M$ the usual $L_2$-space of square integrable functions on $M$ with respect to
the volume measure on $M$. Locally in a chart, the volume measure has the
density $(\det g)^\frac12$ with respect to the Lebesgue measure, where $\det
g$ is the determinant of the metric tensor $(g_{ij})$ in this chart. The norm
of $\Lsqr M$ will be denoted by $\norm[M] \cdot$. For $u \in \Cci M$, the
space of compactly supported smooth functions, we set
\begin{displaymath}
  \check q_M(u):=\int_{M} |\dd u|^2.
\end{displaymath}
\sloppy Here $\dd u$ denotes the exterior derivate of $u$, which is a section
in the cotangent bundle $T^* \! M$ over $M$. Locally in a chart, $| \dd u |^2=
\sum_{i,j} g^{ij} \partial_i u \, \partial_j \overline u$ where $(g^{ij})$ is
the inverse of $(g_{ij})$.

We denote the closure of the non-negative quadratic form $\check q_M$ by
$q_M$.  Note that the domain $\dom q_M$ of the closed quadratic form $q_M$
consists of those functions in $L_2(M)$ such that the weak derivative $\dd u$
is also square integrable (i.e., $q_M(u) < \infty$).

We define the \emph{Laplacian} $\laplacian M$ (for a manifold without
boundary) as the unique self-adjoint and non-negative operator associated with
the closed quadratic form $q_M$, i.e., operator and quadratic form are related
by
\begin{displaymath}
  q_M(u)=\iprod {\laplacian M u} u
\end{displaymath}
for $u \in \Cci M$ (for details on quadratic forms see
e.g.~\cite[Chapter~VI]{kato:66}, \cite{reed-simon-1} or \cite{davies:96}).

If $M$ is a compact manifold with (piecewise) smooth boundary $\bd M \ne
\emptyset$ we can define the Laplacian with \emph{Dirichlet} resp.\ 
\emph{Neumann boundary conditions} in the same way. Here, we start from the
(closure of the) quadratic form $\check q_M$ defined on $\Cci M$, the space of
smooth functions with support \emph{away} from the boundary, resp.\ on $\Ci
M$, the space of functions smooth up to the boundary. We denote the closure of
the quadratic form by $q_M^\Dir$ resp.\ $q_M^\Neu$ and the corresponding
operator by $\laplacianD M$ resp.\ $\laplacianN M$.

If $M$ is compact the spectrum of $\laplacian M$ (with any boundary condition
if $\bd M \ne \emptyset$) is purely discrete. We denote the corresponding
eigenvalues by $\EW k M$ (resp.\ $\EWD k M$ or $\EWN k M$ in the Dirichlet or
Neumann case) written in increasing order and repeated according to
multiplicity.

\subsection*{Periodic manifolds and Floquet Theory}

Let $\Gamma$ be an abelian group of infinite order with $r$ generators and
neutral element $1$. Such groups are isomorphic to $\Z^{r_0} \times
\Z_{p_1}^{r_1} \times \dots \times \Z_{p_a}^{r_a}$ with $r_0>0$ and $r_0 +
\dots + r_a=r$.  Here, $\Z_p$ denotes the abelian group of order $p$.  A
$d$-dimensional (non-compact) Riemannian manifold $\Mper$ will be called
\emph{$\Gamma$-periodic} or a \emph{covering manifold} if $\Gamma$ acts
properly discontinuously, isometrically and cocompactly, i.e., the quotient
$\Mper / \Gamma$ is a $d$-dimensional compact Riemannian manifold such that
the quotient map is a local isometry (cf.~e.g.~\cite{atiyah:76, bruening:92,
  chavel:93, donnelly:81, sunada:89}). For simplicity we assume that $\Mper$
has no boundary.

A compact subset $M$ of $\Mper$ is called a \emph{period cell} if $M$ is the
closure of a fundamental domain $D$, i.e., $M=\overline D$, $D$ is open and
connected, $D$ is disjoint from any translate $\gamma D$ for all $\gamma \in
\Gamma$, $\gamma \ne 1$, and the union of all translates $\gamma M$ is equal
to $\Mper$.

Now we want to analyse the spectrum of periodic elliptic operators on $\Mper$.
Here, \emph{periodicity} means that the operator commutes with all translation
operators on $\Lsqr \Mper$ induced by the group action
(cf.~e.g.~\cite{bruening:92, sunada:89}). In particular the
Laplacian on $\Mper$ is periodic. From Floquet theory it suffices to analyse
the spectra of the periodic operator restricted to a period cell $M$ with
quasi-periodic boundary conditions (cf.~e.g.~\cite{donnelly:81,
  reed-simon-4}).  In order to do this, we define $\theta$-periodic boundary
conditions. Let $\theta$ be an element of the dual group $\hat \Gamma =
\Hom(\Gamma,\Torus^1)$ of $\Gamma$, which is isomorphic to a subgroup of the
$r$-dimensional torus $\Torus^r = \{ \theta \in \C^r; \, |\theta_i|=1 \text{
  for all $i$} \}$. We denote by $q_M^\theta$ the closure of the quadratic
form $\check q_M$ defined on the space of those functions $u \in \Ci M$ that
satisfy
\begin{displaymath}
  u(\gamma x)= \overline {\theta(\gamma)} \, u(x)
\end{displaymath}
for all $x \in \bd M$ and all $\gamma \in \Gamma$ such that $\gamma x \in \bd
M$. The corresponding operator is denoted by $\laplacianT M$. Again,
$\laplacianT M$ has purely discrete spectrum denoted by $\EWT k M$. The
eigenvalues depend continuously on $\theta$. Furthermore $\EWT k M$ depends
even analytically on $\theta$ if we exclude those $\theta \in \hat \Gamma$ for
which $\EWT k M$ is a multiple eigenvalue (cf.~e.g.~\cite{bratteli:99,
  reed-simon-4}).  From Floquet theory we obtain
\begin{displaymath}
  \spec \laplacian \Mper = \bigcup_{\theta \in \hat \Gamma} \spec \laplacianT M
                         = \bigcup_{k \in \N} B_k(\Mper)
\end{displaymath}
where $B_k=B_k(\Mper) = \{ \EWT k M ; \, \theta \in \hat \Gamma\}$ is a
compact interval, called \emph{$k$-th band} (cf.~e.g.~\cite{reed-simon-4}).
In general, we do not know whether the intervals $B_k$ overlap or not. But we
can show the existence of gaps by proving that $\EWT k M$ does not vary too
much in $\theta$.

\begin{remark}
  \label{rem:abs.cont}
  Note that the first band cannot be trivial (i.e. $B_1=\{0\}$) since the
  first eigenvalue $\EWT 1 M$ is $0$ if and only if we are in the case of
  periodic boundary conditions, i.e., $\theta=1$. There are no constant
  \emph{and} $\theta$-periodic functions if $M$ is connected. This means that
  the first band consists of absolutely continuous spectrum provided the first
  and second band do not overlap or more precisely, that $B_1 \cap B_2^c$ is
  absolutely continuous. In general it is not true that all the spectrum is
  absolutely continuous. In~\cite{sunada:89} one can find an example where a
  band reduces to a point (not necessarily being an isolated eigenvalue).
\end{remark}

\subsection*{Main Lemma and Min-max principle}
Here we state a formal result on how to deal with parameter-depending Hilbert
spaces and operators with purely discrete spectrum on these spaces. In
particular, we are interested in the dependence of the eigenvalues on the
parameter. Such Hilbert spaces occur in the next section when we construct a
period cell depending on a parameter $\eps$. The basic idea of the Main
Lemma~\ref{lem:main.lemma} is taken from \cite{anne:87} and \cite{fukaya:87}.
Nevertheless the Main Lemma is more general: here we allow even non-uniform
convergence, i.e., the convergence assumed in Conditions~\eqref{eq:cond.norm}
and~\eqref{eq:cond.quad} could depend on $(u_\eps)$. Furthermore, we can
choose alternatively between the convergence or the inequality in the
assumptions.

We first quote the \emph{Min-max Principle}. Suppose that $q$ is a closed,
non-negative qua\-dra\-tic form on the separable Hilbert space $\HS$ such that
the corresponding operator $Q$ has purely discrete spectrum denoted by $\spec
Q = \{\lambda_k \,|\, k \in \N \}$. Note that $\lambda_k \ge 0$ for all $k$.
Throughout this article we assume that the sequence of eigenvalues
$(\lambda_k)$ is written in increasing order and repeated according to
multiplicity. We then have
\begin{equation}
  \label{eq:max.min}
  \lambda_k = \inf_{L_k} \sup_{u \in L_k, u \ne 0} \frac {q(u)}{\normsqr u}
\end{equation}
where the infimum is taken over all $k$-dimensional subspaces $L_k$ of $\dom
q$ (for this version of the Min-max principle see e.g.~\cite{davies:96}).

Suppose now that for each $\eps>0$ we have separable Hilbert spaces $\HS_\eps$
and $\HS'_\eps$. Furthermore suppose that $q_\eps$ and $q'_\eps$ are
non-negative, closed quadratic forms on $\HS_\eps$ and $\HS'_\eps$. Finally
suppose that the corresponding operators have purely discrete spectrum
denoted by $\lambda_k(\eps)$ and $\lambda'_k(\eps)$, $k \in \N$ (written in
increasing order and repeated according to multiplicity). A corresponding
orthonormal basis of eigenfunctions of $Q_\eps$ is denoted by
$(\phi_k^\eps)_k$ and the linear span of the first $k$ eigenvalues by
$L_k(\eps)$.
\begin{lemma}[Main Lemma]
  \label{lem:main.lemma}
  Suppose that for each $\eps>0$ a linear map $\Phi_\eps \colon \dom q_\eps
  \longrightarrow \dom q'_\eps$ is given such that for all $u_\eps \in
  L_k(\eps)$ Conditions~\eqref{eq:cond.norm} and~\eqref{eq:cond.quad} are
  satisfied:
  \begin{align}
  \label{eq:cond.norm}
    \lim_{\eps \to 0} ( \normsqr[\HS'_\eps]{\Phi_\eps u_\eps}  -
                        \normsqr[\HS _\eps]{          u_\eps}) &= 0
    & \text{or} &&
                        \normsqr[\HS _\eps]{          u_\eps}  &\le
                        \normsqr[\HS'_\eps]{\Phi_\eps u_\eps}  \\
  \label{eq:cond.quad}
    \lim_{\eps \to 0} ( q'_\eps(\Phi_\eps u_\eps) -
                        q _\eps(          u_\eps) ) &= 0
    & \text{or} &&
                        q _\eps(          u_\eps)   &\ge
                        q'_\eps(\Phi_\eps u_\eps).
  \end{align}
  Furthermore, we assume that for each $k \in \N$ there exist a constant
  $c_k>0$ such that
  \begin{equation}
  \label{eq:cond.ew}
    \lambda_k(\eps) \le c_k \qquad \text{for all $\eps>0$.}
  \end{equation}
  Then for each $k \in \N$ there exists a function $\delta_k(\eps) \ge 0$
  converging to $0$ as $\eps \to 0$ such that
  \begin{equation}
    \label{eq:ew.est}
    \lambda'_k(\eps) \le \lambda_k(\eps) + \delta_k(\eps)
  \end{equation}
  for small enough $\eps > 0$.
\end{lemma}
\begin{proof}
  For $u=u_\eps = \sum_{i=1}^k \alpha_i^\eps \phi_i^\eps$ with complex numbers
  $\alpha_i=\alpha_i^\eps$ we have
  \begin{displaymath}
    \frac{q'_\eps(\Phi_\eps u)} {\normsqr{\Phi_\eps u}} -
    \frac{q _\eps(          u)} {\normsqr{          u}}    =
    \frac 1 {\normsqr{\Phi_\eps u}}
    \Bigl(
      \frac{q _\eps(          u)} {\normsqr{        u}}
      \bigl(
        \normsqr{u} - \normsqr{\Phi_\eps u}
      \bigr) +
      \bigl(
        q'_\eps(\Phi_\eps u) - q_\eps(u)
      \bigr) 
    \Bigr).
  \end{displaymath}
  Furthermore, we estimate
  \begin{multline}
  \label{eq:est.norm}
     \normsqr{u} - \normsqr{\Phi_\eps u} =
     \sum_{i,j=1}^k \alpha_i \overline {\alpha_j}
     \bigl(
       \delta_{ij} - \iprod {\Phi_\eps \phi_i^\eps} {\Phi_\eps \phi_j^\eps}
     \bigr) \\ \le
     \delta'_k(\eps) \sum_{j=1}^k |\alpha_j|^2 = 
     \delta'_k(\eps) \, \normsqr {u}
  \end{multline}
  where
  \begin{displaymath}
    \delta'_k(\eps) := k \max_{i,j = 1, \dots, k} 
      |\delta_{ij} - \iprod {\Phi_\eps \phi_i^\eps } {\Phi_\eps \phi_j^\eps }|
  \end{displaymath}
  by the Cauchy-Schwarz Inequality. The Polarisation Identity
  together with Condition~\eqref{eq:cond.norm} yields $\delta'_k(\eps) \to 0$
  as $\eps \to 0$. If we are in the second alternative of
  Condition~\eqref{eq:cond.norm} we simply set $\delta'_k(\eps)=0$.  By a
  similar argument we can show the existence of a function $\delta''_k(\eps)
  \ge 0$ converging to $0$ as $\eps \to 0$ such that
  \begin{equation}
  \label{eq:eq:est.quad}
     q'_\eps(\Phi_\eps u) - q_\eps(u) \le
     \delta''_k(\eps) \normsqr{u}.
  \end{equation}
  From Estimate~\eqref{eq:est.norm} we also conclude
  \begin{equation}
  \label{eq:est.norm2}
    \normsqr{u} \le \frac 1 {1-\delta'_k(\eps)} \normsqr{\Phi_\eps u}
  \end{equation}
  provided $\eps$ is small enough. Applying Condition~\eqref{eq:cond.ew} we
  obtain the estimate
  \begin{equation}
  \label{eq:est.ew}
    \frac{q'_\eps(\Phi_\eps u)} {\normsqr{\Phi_\eps u}} -
    \frac{q _\eps(          u)} {\normsqr{          u}}    \le
    \delta_k(\eps):= \frac 1 {1-\delta'(\eps)} 
      \bigl( 
         c_k \delta'(\eps) + \delta''(\eps)
      \bigr).
  \end{equation}
  Estimate~\eqref{eq:est.norm2} also yields the injectivity of
  $\Phi_\eps \restr {L_k(\eps)}$, i.e., $\Phi_\eps(L_k(\eps))$ is a
  $k$-dimen\-sional subspace of $\dom q'_\eps$. Finally, the Min-max
  Principle~\eqref{eq:max.min} implies the desired estimate on the
  eigenvalues.
\end{proof}

\section{Construction of a Periodic Manifold}
\label{sec:constr}

Suppose that $X$ is a compact oriented and connected Riemannian manifold of
dimension $d \ge 2$ (for simplicity without boundary). We want to construct a
$\Gamma$-periodic manifold where $\Gamma$ is an abelian group with $r$
generators $e_1, \dots, e_r$. We choose $2r$ distinct points
$x_1,\dots,x_{2r}$. For each point $x_i$, denote by $B^i_\eps$ the open
geodesic ball around $x_i$ of radius $\eps>0$.  Suppose further that
$B^i_{\eps_0}$ are pairwise disjoint where $\eps_0>0$ denotes the injectivity
radius of $X$. Denote by $B_\eps$ the union of all balls $B^i_\eps$, $i=1,
\dots, 2r$. Let $\Xeps := X \setminus B_{2\eps}$ for $0 < 2\eps < \eps_0$ with
metric inherited from $X$.

\begin{figure}[ht]
  \begin{center}
    \leavevmode
    \begin{picture}(0,0)(-8,0)
      \includegraphics{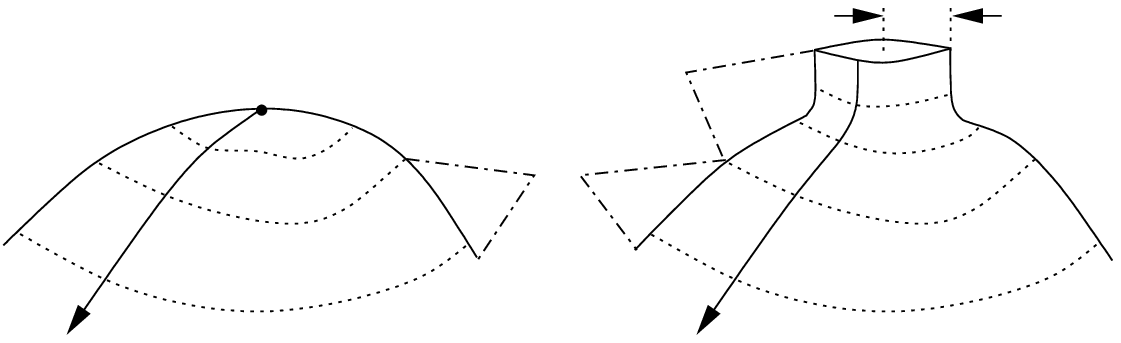}
    \end{picture}
    \setlength{\unitlength}{4144sp}
    \begin{picture}(5094,1654)(195,-884)
      \put(4816,-826){$\Meps$}
      \put(4341,614){$\eps$}
      \put(1801,-826){$X$}
      \put(1261,299){$x_i$}
      \put(361,-781){$s$}
      \put(3196,-781){$s$}
      \put(521,-421){$\eps_0$}
      \put(3411,-421){$\eps_0$}
      \put(856,-71){$2\eps$}
      \put(3721,-66){$2\eps$}
      \put(3946,149){$\eps$}
      \put(1101,104){$\eps$}
      \put(4146,299){$\eps/2$}
      \put(2639,-115){$\Xeps$}
      \put(3129,365){$A_\eps^i$}
    \end{picture}
    \caption{Modification of the metric $g$ of the $2$-dimensional manifold
      $X$ near the point $x_i$.}
    \label{fig:gaps2}
  \end{center}
\end{figure}

On $B^i_{\eps_0}$, the metric of $X$ is given in polar coordinates
$(s,\sigma) \in ]0,\eps_0[ \times \Sphere^{d-1}$ by
\begin{equation}
  \label{eq:met.pol.coord}
  g = \dd s^2 + h^i_s
\end{equation}
where $h^i_s$ denotes a metric on $\{s\} \times \Sphere^{d-1}$ (see
Figure~\ref{fig:gaps2}). Here, $\Sphere^{d-1}$ denotes the $(d-1)$-dimensional
sphere with standard metric $\dd \sigma^2$.

Let $r_\eps$ be a smooth monotone function with $r_\eps(s)=\eps$ for
$0 \le s \le \eps/2$ and $r_\eps(s)=s$ for $2\eps \le s \le \eps_0$.
Furthermore, let $\chi_\eps$ be a smooth cut-off function having values
between $0$ and $1$ and satifying $\chi_\eps(s)=0$ if $s \le \eps$ and
$\chi_\eps(s)=1$ if $s \ge 2\eps$. Now we let the modified metric
$h^i_{\eps,s}$ be a convex combination of the original metric and the
spherical metric $(r_\eps(s))^2 \dd \sigma^2$, i.e.,
\begin{equation}
  \label{eq:conv.comb}
  h^i_{\eps,s} := 
  \chi_\eps(s) h^i_s + \bigl( 1-\chi_\eps(s) \bigr) 
                       \bigl( r_\eps(s) \bigr)^2 \dd \sigma^2.
\end{equation}
We denote the completion of $X \setminus \{x_1, \dots, x_{2r}\}$ together with
the modified metric
\begin{displaymath}
  g^i_\eps := \dd s^2 + h^i_{\eps,s} 
\end{displaymath}
on $B^i_{\eps_0}$ by $\Meps$ (see Figure~\ref{fig:gaps2}).  Since
$g^i_\eps(s,\sigma) = g(s,\sigma)$ for $s \ge 2 \eps$, the punched manifold
$\Xeps$ is embedded in $\Meps$. Furthermore, since $g^i_\eps(s,\sigma) = \dd
s^2 + \eps^2 \dd \sigma^2$ for $s \le \eps/2$, there exists a neighbourhood of
$\bd \Meps \cap B^i_\eps$ given in coordinates by $[0,\eps/2] \times
\Sphere^{d-1}$ which is isometric to a cylinders of radius $\eps$ and length
$\eps/2$.
Let $A^i_\eps$ be the \emph{cylindrical end} of the manifold $\Meps$ near
$x_i$ given in coordinates by $[0,2\eps] \times \Sphere^{d-1}$. Next, let
$A^i_{\eps,x}$ be the sphere with distance $s$ from the boundary given in
coordinates by $\{s\} \times \Sphere^{d-1}$.  Finally, let $\Aeps$ be the
union of all cylindrical ends $A^i_\eps$, $i = 1, \dots, 2r$.

Finally, we construct the corresponding periodic manifold $\Mepsper$: Let
$\gamma \Meps$ be an isometric copy of $\Meps$ with identification $x \mapsto
\gamma x$ for each $\gamma \in \Gamma$. We construct a new (non-compact)
manifold $\Mepsper$ by identifying $\gamma \bd A^{2i-1}_\eps$ with $e_i \gamma
\bd A^{2i}_\eps$ for each $\gamma \in \Gamma$ and $i=1, \dots, r$. Remember
that $e_i$ denotes the $i$-th generator of $\Gamma$. Since in a neighbourhood
of $\bd A^i_\eps$ the manifold is isometric to a cylinder of radius $\eps$, we
can choose a smooth atlas and a smooth metric on the glued manifold
$\Mepsper$. We therefore obtain a (non-compact) $\Gamma$-periodic manifold
$\Mepsper$ and $\Meps$ is a period cell for $\Mepsper$.

Now we are able to state the following theorem (Theorem~\ref{thm:gaps.constr}
follows via Floquet Theory):
\begin{theorem}
  \label{thm:conv.constr}
  As $\eps \to 0$ we have $\EWT k \Meps \to \EW k X$ uniformly in $\theta \in
  \hat \Gamma$.
\end{theorem}
Therefore, the $k$-th band $B_k(\Mepsper)$ reduces to the point $\{ \EW k X
\}$ as $\eps \to 0$.  Note that the convergence is \emph{not} uniform in $k$
since there are topological obstructions (see the discussion in
\cite{chavel-feldman:81}). We therefore could not expect that an
\emph{infinite} number of gaps occur.

Before we prove Theorem~\ref{thm:conv.constr}, we need two lemmas. The idea
is to compare the $\theta$-periodic eigenvalues on $\Meps$ with Dirichlet and
Neumann eigenvalues on $\Xeps$. The crucial point is, that the corresponding
$\theta$-periodic eigenfunctions on $\Meps$ do not concentrate on $\Aeps$,
i.e., on the cylindrical ends. This will be shown in the second lemma. First
we need to compare the density of the $(d-1)$-dimensional volume of
$A^i_{\eps, s}$ with the volume of the sphere of radius $r_\eps(s)$:
\begin{lemma}
  \label{lem:mod.met.est}
  There exists a constant $c \ge 1$ such that
  \begin{displaymath}
    \frac 1c \, r_\eps(s)^{d-1} \le 
    (\det h^i_{\eps,s})^\frac12 \le
    c        \, r_\eps(s)^{d-1}
  \end{displaymath}
  for all $0 \le s \le \eps_0$ and all $i$.
\end{lemma}
\begin{proof}
  The metric $g= \dd s^2 + h^i_s$ on $B^i_{\eps_0}$ can be compared with the
  flat metric $\dd s^2 + s^2 \dd \sigma^2$ (pointwise in the sense of
  sesquilinear forms), i.e., there exists a constant $c'\ge 1$ such that
  \begin{displaymath}
    \frac1{c'} \, s^2 \dd \sigma \le
    h^i_s \le
    c'         \, s^2 \dd \sigma.
  \end{displaymath}
  By our assumptions on $r_\eps$ the same estimate is true with $s^2$ replaced
  by $r_\eps(s)^2$ and $h^i_s$ replaced by the convexe
  combination~\eqref{eq:conv.comb}. The result follows from the monotonicity
  of $\det$.
\end{proof}

Now we prove the non-concentration of the eigenfunctions on the cylindrical
ends as $\eps \to 0$:
\begin{lemma}
  \label{lem:no.conc}
  There exists a positive function $\omega(\eps)$ converging to $0$ as $\eps
  \to 0$ such that
  \begin{equation}
    \label{eq:no.conc}
    \int_{\Aeps} |u|^2 \le
    \omega(\eps) \int_{\Meps} \bigl( |u|^2 + | \dd u|^2 \bigr),
  \end{equation}
  for all $u$ out of the domain of the quadratic form with
  $\theta$-periodic boundary conditions on $\Meps$.
\end{lemma}
Note that $\omega(\eps)$ only depends on the geometry of $X$ near $x_i$.
\begin{proof}
  Without loss of generality, we can assume that $u \in \Ci \Meps$. Suppose
  furthermore that $u(\eps_0,\sigma)=0$ for all $\sigma \in \Sphere^{d-1}$.
  First we show an $L_2$-estimate over $A^i_{\eps,s}$ with its induced metric
  $h^i_{\eps,s}$.
  
  Applying the Cauchy-Schwarz Inequality and Lemma~\ref{lem:mod.met.est}
  yields
  \begin{multline*}
    |u(s,\sigma)|^2 =
    \Big| \int_s^{\eps_0} \partial_t u(t,\sigma)\, \dd t\Big|^2 \\ \le
       c \int_s^{\eps_0} \hspace{-1ex} r_\eps(t)^{1-d}  \dd t \, 
         \int_s^{\eps_0} |\partial_t u (t,\sigma)|^2 
                 (\det h^i_{\eps,t}(\sigma))^\frac12 \dd t.
  \end{multline*}
  If we integrate over $\sigma \in \Sphere^{d-1}$ and apply
  Lemma~\ref{lem:mod.met.est} once more we obtain
  \begin{align}
    \int_{A^i_{\eps,s}} |u|^2  &= 
    \int_{\Sphere^{d-1}} |u(s,\sigma)|^2 
     (\det h^i_{\eps,s}(\sigma))^\frac12 \nonumber\\ 
     \label{eq:int.est}  & \le
     c^2 r_\eps(s)^{d-1} \int_s^{\eps_0} r_\eps(t)^{1-d}  \dd t
                     \int_{\Meps} |\dd u|_{T^* \! \Meps}^2.
  \end{align}
  If $0 \le s \le 2\eps$ we have $r(s)^{d-1} \le (2\eps)^{d-1}$. Furthermore,
  the integral over $t$ can be split into an integral over $0 \le t \le 2\eps$
  and $2\eps \le t \le \eps_0$. The first integral can be estimated by
  $\eps^{2-d}$, the second by $\int_{2\eps}^{\eps_0} t^{1-d} \dd t$. Therefore
  we have an estimate of the order $O(\eps)$ if $d \ge 3$ resp.\ $O(\eps|\ln
  \eps|)$ if $d=2$. Finally, if we integrate the integral on the LHS
  of~\eqref{eq:int.est} over $s \in [0, 2\eps]$ we obtain the desired
  Estimate~\eqref{eq:no.conc}.
  If $u(\eps_0,\sigma)\ne 0$ we choose a cut-off function.
\end{proof}

The argument in the proof is due to \cite{anne:87}. The following lemma is
proven in \cite{chavel-feldman:78} resp.\ \cite{anne:87}.

\begin{lemma}
  \label{lem:anne.chavel}
  We have $\EWD k \Xeps \to \EW k X$ resp.\ $\EWN k \Xeps \to \EW k X$.
\end{lemma}

Now we can show Theorem~\ref{thm:conv.constr}:
\begin{proof}
  From the Min-max Principle~\eqref{eq:max.min} we conclude
  \begin{equation}
    \label{eq:upper.est}
    \EWT k \Meps \le  \EWD k \Xeps
  \end{equation}
  since the domains of the quadratic forms obey the opposite inclusions.  In
  particular, $\EWT k \Meps$ is bounded in $\theta$ and $\eps$ by some
  constant $c_k>0$. To prove the opposite inequality we apply our Main
  Lemma~\ref{lem:main.lemma} with $\HS_\eps := \Lsqr \Meps$, $q_\eps :=
  q^\theta_{\Meps}$, $\HS'_\eps := \Lsqr \Xeps$, $q'_\eps := q^\Neu_{\Xeps}$
  and $\Phi_\eps u := u \restr \Xeps$ being the restriction operator.
  
  \sloppy
  Condition~\eqref{eq:cond.norm} is satisfied by Lemma~\ref{lem:no.conc},
  the inequality in Condition~\eqref{eq:cond.quad} is trivially
  satisfied. Finally, Condition~\eqref{eq:cond.ew} is satisfied by the upper
  bound~\eqref{eq:upper.est} and Lemma~\ref{lem:anne.chavel}. The Main Lemma
  therefore yields
  \begin{displaymath}
    \EWN k \Xeps \le \EWT k \Meps + \delta_k(\eps)
  \end{displaymath}
  with $\delta_k(\eps) \to 0$ as $\eps \to 0$. Together with
  Estimate~\eqref{eq:upper.est} and Lemma~\ref{lem:anne.chavel} we are done.
\end{proof}

\begin{remark}
  \label{rem:constr.curv}
  If we assume that the metric $g$ of the original manifold $X$ is flat on
  $B^i_{\eps_0}$, i.e., $h^i_s=s^2 \dd \sigma^2$ in
  Equation~\eqref{eq:met.pol.coord}, we can calculate the sectional curvature
  of the cylindrical end $\Aeps$ of $\Meps$.  Let $\partial_s$,
  $\partial_{\sigma_1}, \dots, \partial_{\sigma_{d-1}}$ be orthonormal tangent
  vectors corresponding to $(s,\sigma)$ on $A^i_\eps$.  Then the sectional
  curvature is
    \begin{displaymath}
      K(\partial_s,\partial_{\sigma_j}) = -\frac{\ddot r_\eps} {r_\eps}
          \qquad \text{and} \qquad
      K(\partial_{\sigma_j},\partial_{\sigma_k}) = 
         \frac{1 - \dot r_\eps^2} { r_\eps^2}
    \end{displaymath}
    for $j \ne k$. With our assumptions on $r_\eps$,
    $K(\partial_s,\partial_{\sigma_j})$ is a negative number of order
    $O(\eps^{-2})$ and $K(\partial_{\sigma_j},\partial_{\sigma_k})$ is a
    positive number of the same orderas $\eps \to 0$. In particular, the Ricci
    and the scalar curvature are also of order $O(\eps^{-2})$. Therefore, all
    curvature terms are unbounded in $\eps$ as Green conjectured in
    \cite{green:97}.
\end{remark}

\subsection*{Periodic manifold joined by cylinders}
In the previous example $\Mepsper$ one may think that it is essential for
obtaining spectral gaps that the parts which break down reduce to a point. To
avoid this impression we will confront another periodic manifold with spectral
gaps where the collapsing parts reduce to intervals instead of points. It only
seems to be important that the period cells are seperated by very short closed
geodesics (in dimension $2$) or more generally by very small submanifolds of
codimension $1$. The construction in this section is closely related to the
work of Ann\'e~\cite{anne:87}.

For simplicitywe will only assume that $\Gamma=\Z$. Let $I$ be the interval
$[0,L]$ and let $\Ceps$ be the cylinder $I \times \Sphere^{d-1}$ of radius
$\eps$ and length $L>0$ with metric $\dd s^2 + \eps^2 \dd \sigma^2$ (if $L=0$
we are in the case of the previous section) . We denote by $\Neps$ the period
cell $\Meps$ where we have glued the cylinder $\Ceps$ by identifying $\{0\}
\times \Sphere^{d-1}$ with $\bd A^1_\eps$. In the same way as above we obtain
a periodic manifold $\Nepsper$ with period cell $\Neps$ by glueing together
$\Z$ copies of $\Neps$. As above we can prove:
\begin{theorem}
  \label{thm:conv.cyl}
  As $\eps \to 0$ we have $\EWT k \Neps \to \EWD k {X \dcup I}$ uniformly in
  $\theta \in \hat \Gamma$.
\end{theorem}
Here, $\EWD k {X \dcup I}$ denotes the eigenvalues of the operator $\laplacian
X \oplus \laplacianD I$ written in increasing order and repeated according to
multiplicity. Note that $\spec (\laplacian X \oplus \laplacianD I) = \spec
\laplacian X \cup \spec \laplacian I$, i.e., $\EWD k {X \dcup I}$ is a
reordering of $\EWD {k_1} X$ and $\EWD {k_2} I$. Again, by Floquet Theory
Theorem~\ref{thm:gaps.cyl} follows.

\begin{proof}
  We only sketch the proof since it is similar to the proof of
  Theorem~\ref{thm:conv.constr}. Again, we have an upper estimate of $\EWT k
  \Neps$ by the $k$-th Dirichlet eigenvalue $\EWD k {\Xeps \dcup \Ceps}$ on
  $\Xeps \dcup \Ceps$. For the estimate from below we apply the Main Lemma
  once more with $\HS_\eps := \Lsqr \Neps$, $q_\eps := q^\theta_{\Neps}$,
  $\HS'_\eps := \Lsqr \Xeps \oplus \Lsqr \Ceps$ and $q'_\eps := q^\Neu_{\Xeps}
  \oplus q^\Dir_{\Ceps}$. Furthermore, for $u \in \dom q^\theta_{\Neps}$, we
  set
  \begin{displaymath}
    \Phi_\eps u := 
    u \restr \Xeps \oplus (u \restr \Ceps -h)   
  \end{displaymath}
  where $h=h_\eps$ is the (unique) function satisfying
  \begin{align*}
    \laplacian \Ceps h &= 0 &\text{and}&&
    h \restr {\bd \Ceps} &= u \restr {\bd \Ceps}
  \end{align*}
  (see~\cite{anne:87} and \cite{post:00}).  To verify
  Condition~\ref{eq:cond.norm}, we estimate
  \begin{multline*}
    \bigl| \normsqr[\Xeps \dcup \Ceps]{\Phi_\eps u} - 
           \normsqr[\Neps] u 
    \bigr| \le
    \int_{\Aeps} \hspace{-1ex} |u|^2 + 
    \int_{\Ceps} \hspace{-1ex} \bigl( |u-h|^2 - |u|^2 \bigr) \\ \le
    \normsqr[\Aeps] u + 2 \norm[\Ceps] u \norm[\Ceps] h + \normsqr[\Ceps] h.
  \end{multline*}
  By Lemma~\ref{lem:no.conc} we only need to show that the harmonic extenstion
  $h$ converges to $0$ in $\Lsqr \Ceps$ uniformly in $\theta$ (note that
  $h=h^\theta_\eps$ depends also on $\theta$ since $u=u^\theta_\eps \in \dom
  q^\theta_\Meps$ does). The convergence does not seem to be very surprising
  since the harmonic function $h$ is given on the boundary $\bd \Ceps$ by $u$
  which is small (in $L_2$-sense) by Lemma~\ref{lem:no.conc}. Nevertheless a
  little more work is necessary which will be omitted here (see~\cite{anne:87}
  and \cite{post:00}).

  Condition~\ref{eq:cond.quad} is satisfied since
  \begin{displaymath}
    q_\eps(u) = 
   \int_\Neps | \dd u|^2 \ge 
   \int_\Xeps | \dd u|^2 + \int_{\Ceps} | \dd h|^2 =
    q'_\eps(\Phi_\eps u).
  \end{displaymath}
  Note that the harmonic function $h$ minimizes the energy integral
  $q_\Ceps(u)$.  Since 
  \begin{equation}
    \label{eq:harmonic}
   \int_\Ceps \iprod {\dd (u-h)} {\dd h} = 0 
  \end{equation}
  by the Gauss-Green Formula we have $q_\Ceps(u) = q_\Ceps(h)+q_\Ceps(u-h) \ge
  q_\Ceps(h)$. Note that $u-h$ satisfies Dirichlet boundary conditions.
  
  From the Main Lemma we obtain a lower bound (up to an error term) given by
  the $k$-th eigenvalue $\EWND k {\Xeps \dcup \Ceps}$ with Neumann boundary
  conditions on $\Xeps$ and Dirichlet boundary conditions on $\Ceps$. We only
  have to add on that $\EWND k {\Xeps \dcup \Ceps}$ resp.\ $\EWD k {\Xeps
    \dcup \Ceps}$ converge to $\EWD k {X \dcup I}$ if the cylinder $\Ceps$
  collapes to the interval $I$ as $\eps \to 0$.
\end{proof}

\section{Conformal Deformation}
\label{sec:conf}

Suppose $\Mper$ is a $\Gamma$-periodic Riemannian manifold of dimension $d \ge
2$ with metric $g$. Let $X \subset \Mper$ be a compact subset with smooth
boundary such that $\gamma X \cap X \ne \emptyset$ implies $\gamma=0$. Then a
period cell $M$ with $\dist(\bd X, \bd M)>0$ exists.  We introduce
\emph{normal} or \emph{Fermi coordinates} $(r,y)$ with respect to $Y:=\bd X$
(for details cf.\ \cite{chavel:93}). Here, $r \in (-r_0,r_0)$ parametrises
the normal direction and $y \in Y$ parametrises the tangential direction;
$r<0$ corresponds to the interior of $X$ and $r=0$ corresponds to $Y$.

Furthermore, we assume that normal coordinates also exist on $\overline{M
  \setminus X}$, i.e., we suppose that $\overline{M \setminus X}$ can be
parametrised by $(r,y)$ with $r \in I_y$ and $y \in Y$. Here, $I_y$ is a
compact subset of $\R$ containing $[0,r_0]$.  The existence of normal
coordinates on $\overline{M \setminus X}$ is a geometrical restriction on $X$.
In some situations this condition means that $X$ is close enough to $\bd M$.
For example, this condition is satisfied for a centered ball in a cube.

Suppose we have for each $\eps>0$ a smooth $\Gamma$-periodic function
$\rho_\eps \colon \Mper \longrightarrow (0,1]$ with the following
properties:
\begin{align}
  \label{eq:cond.one}
  \rho_\eps(x) &=1    
               &&\text{for all $x \in X$,} \\
  \label{eq:cond.small}
  \rho_\eps(x) &=\eps 
               &&\text{for all $x \in M$ with $\dist(x,X) \ge \eps^d$.}
\end{align}
For simplicity, we also assume that $\rho_\eps(x)$ is only a function on $r$
in normal coordinates on $\overline{M \setminus X}$.  Note that the function
$\rho_\eps$ converges pointwise to the indicator function of the set
$\Xper=\bigcup_{\gamma \in \Gamma} X$. We define $g_\eps := \rho_\eps^2 g$ and
denote the resulting Riemannian manifolds with metric $g_\eps$ by $\Mepsper$
resp.\ $\Meps$. We therefore obtain a conformally deformed $\Gamma$-periodic
manifold $\Mepsper$ with periodic metric $g_\eps$ and period cell $\Meps$.  In
particular, the squared norm and the quadratic form on the deformed manifold
$\Meps$ are given by
\begin{align}
  \label{eq:conf.fact}
  \normsqr[\Meps] u &= \int_M |u|^2 \rho_\eps^d       
                    &\text{and}&&
  q_\Meps (u)       &= \int_M |\dd u|_{T^*\!M}^2 \,
                                        \rho_\eps^{d-2}.
\end{align}
Here we can see that the case $d=2$ is in some sense particular, since the
quadratic form does not depend on $\eps$ any more (but the norm does).

First, let us calculate the curvature of the conformally deformed case:
\begin{remark}
  \label{rem:conf.curv}
  We denote the sectional curvatures of $\Meps$ by the subscript $\eps$ and
  the sectional curvatures of $M$ without subscript. Let $\partial_r$,
  $\partial_{y_1}, \dots, \partial_{y_{d-1}}$ be orthonormal basis tangent
  vectors corresponding to the coordinates $(r,y)$ near $Y=\bd X$.  Then the
  sectional curvatures are given by
    \begin{align*}
      K_\eps(\partial_r,\partial_{y_j}) & = 
       \rho_\eps^5 
       \bigl(
           -\ddot\rho_\eps + \partial_r g_{y_j y_j}\dot \rho_\eps/2
           + \rho_\eps K(\partial_r,\partial_{y_j}) 
       \bigr)\\
      K_\eps(\partial_{y_j},\partial_{y_k}) & = 
       \rho_\eps^6
       \bigl( 
            -\dot \rho_\eps^2 - 
               (\partial_r g_{y_j y_j} + \partial_r g_{y_k y_k})
                   \dot \rho_\eps \rho_\eps/2 +
            \rho_\eps^2 K(\partial_{y_j},\partial_{y_k})
       \bigr)
    \end{align*}
    for $j \ne k$. With our assumptions on $\rho_\eps$,
    $K(\partial_r,\partial_{y_j}) = O(\eps^{-2d-5})$ with changing sign and
    $K(\partial_{\sigma_j},\partial_{\sigma_k})=O(\eps^{-2d-6})$ provided
    $\eps > 0$ is small enough. Calculating the Ricci and scalar curvature one
    can see that all curvature terms are not bounded in $\eps$ as Green
    conjectured in \cite{green:97}.
\end{remark}
\subsection*{The Higher Dimensional Case}

Now we are able to state the following theorem (again,
Theorem~\ref{thm:gaps.conf} follows via Floquet Theory):
\begin{theorem}
  \label{thm:conv.conf}
  Suppose that $d \ge 3$. Then $\EWT k \Meps \to \EWN k X$ as $\eps \to 0$
  uniformly in $\theta \in \hat \Gamma$.
\end{theorem}
Again, the $k$-th band $B_k(\Mepsper)$ reduces to the point $\{ \EWN k X
\}$ as $\eps \to 0$ and the convergence is \emph{not} uniform in $k$.

As in the previous section, we need the following lemma which shows that the
$\theta$-periodic eigenfunctions on $\Meps$ do not concentrate on the
metrically shrunken set $\Meps \setminus X$ as $\eps \to 0$:
\begin{lemma}
  \label{lem:no.conc.conf}
  There exists a positive function $\omega(\eps)$ converging to $0$ as $\eps
  \to 0$ such that
  \begin{equation}
    \label{eq:no.conc.conf}
    \int_{\Meps \setminus X} |u|^2 \le
    \omega(\eps) \int_M 
        \bigl( |u|^2 + | \dd u|_{T^*\!\Meps}^2 \bigr),
  \end{equation}
  for all $u$ out of the domain of the quadratic form with
  $\theta$-periodic boundary conditions on $\Meps$.
\end{lemma}
Again, $\omega(\eps)$ only depends on the geometry of $M \setminus X$.
\begin{proof}
  We proceed in the same way as in the proof of Lemma~\ref{lem:no.conc}. We
  introduce normal coordinates.  For notational simplicity only, we assume
  that $I_y=[0,r_y]$ for some number $r_0 \le r_y$. Suppose that $u \in \Ci
  \Meps$ with $u(r,y)=0$ for all $y \in Y$ and $r \le -r_0$. As in
  \eqref{eq:met.pol.coord} we have the orthogonal splitting
  \begin{displaymath}
    g_\eps = \rho_\eps^2 \, g_\eps = \rho_\eps^2 \, (\dd r^2 + h_r)
  \end{displaymath}
  in normal coordinates where $h_r$ is a parameter-dependent metric on $Y$.
  By the Cauchy-Schwarz Inequality we have
  \begin{multline*}
    |u(s,y)|^2 =
    \Big| 
      \int_{-r_0}^s \partial_r u(r,y)\, \dd r 
    \Big|^2  \\ \le
    \int_{-r_0}^s \bigl( \det g(r,y) \bigr)^{-\frac12} \,\dd r
          \cdot 
    \int_{-r_0}^s \bigl( 
                        |\partial_r u|^2 (\det g)^\frac12 
                  \bigr)(r,y) \, \dd r
  \end{multline*}
  for $0 \le s \le r_y$. Since $Y$ is compact we can estimate the first
  integral by $c>0$. Therefore integrating over $s \in I_y$ and $y \in Y$
  yields
  \begin{multline*}
    \int_{\Meps \setminus X} |u|^2 =
    \int_{y \in Y} \int_{s=0}^{r_y}  
         \bigl( 
               |u|^2  (\det g)^\frac12 \rho_\eps^d 
         \bigr)(s,y)  \,\, \dd s \,\dd y 
            \\ \le
    c \int_{y \in Y} \int_{s=0}^{r_y}  (\det g)^\frac12(s,y) \rho_\eps^d(s)
        \int_{r=-r_0}^s 
           \bigl( 
               | \partial_r u|^2 (\det g)^\frac12 
           \bigr)\!(r,y)\,\, \dd r\,\dd s\,\dd y.
  \end{multline*}
  We can estimate the $s$-dependent terms as follows: for $\eps^d \le s \le
  r_y$ we have $\rho_\eps(s)=\eps$ by Assumption~\eqref{eq:cond.small}.
  Furthermore, there exists a constant $c'>0$ such that $(\det g)^\frac12(s,y)
  \le c'$ for all $y \in Y$ and $0 \le s \le r_y$ since $\overline{M \setminus
    X}$ is compact. Therefore the integral over $0 \le s \le \eps^d$ and
  $\eps^d \le s \le r_y$ can be estimated by $c' \eps^d$. We conclude
  \begin{align*}
    \int_{\Meps \setminus X} |u|^2 
    & \le
    c \, c' \eps^d \int_{y \in Y} \int_{r=-r_0}^{r_y} 
         \bigl( 
            | \partial_r u|^2 (\det g)^\frac12 
         \bigr)\!(r,y)\,\, \dd r\, \dd y 
          \\ 
    & \le
    c \, c' \eps^2 \int_{y \in Y} \int_{r=-r_0}^{r_y} 
         \bigl( 
            | \partial_r u|^2 (\det g)^\frac12 \rho_\eps^{d-2} 
         \bigr)\!(r,y)\,\,\dd r\,\dd y \\
    & \le 
    c \, c' \eps^2 \int_M |\dd u|_{T^*\!\Meps}^2
  \end{align*}
  where we have used $\rho_\eps \ge \eps$ in the second line.
  
  If $u(r,y) \ne 0$ for some $y \in Y$ and $r< -r_0$ we multiply $u$ with a
  cut-off function $\chi$ such that $\chi(r) = 1$ for $r \ge -r_0/2$ and
  $\chi(r) = 0$ for $r \le -r_0$. Note that $\supp \chi \subset X$, i.e., on
  $\supp \chi$, there is no conformal deformation. If $u \in \dom
  q^\theta_\Meps$ we apply an approximation argument.
\end{proof}

In the same way we have proven Theorem~\ref{thm:conv.constr} we can show
Theorem~\ref{thm:conv.conf}:
\begin{proof}
  First, we prove an upper bound. For this we apply the Main
  Lemma~\ref{lem:main.lemma} with $\HS := \Lsqr X$, $q := q^\Neu_X$ (not
  depending on $\eps$), $\HS'_\eps := \Lsqr \Meps$ and $q'_\eps :=
  q^\theta_\Meps$. Furthermore, let $\Phi u$ be an extension of $u \in \dom
  q^\Neu_X$ onto $M$ such that $u=0$ in a neighbourhood of $\bd M$. In
  particular, $u \in \dom q^\theta_\Meps$. The inequality of
  Condition~\eqref{eq:cond.norm} is trivially satisfied,
  Condition~\eqref{eq:cond.quad} follows because of
  \begin{displaymath}
    \bigl| q'_\eps(\Phi u) - q(u) \bigr| =
    \int_{\Meps \setminus X} |\dd \Phi u|_{T^*\!\Meps}^2 =
    \int_{\Meps \setminus X} |\dd \Phi u|_{T^*\!\Meps}^2 \rho_\eps^{d-2} \,
                                                     \to 0
  \end{displaymath}
  by the Lebesgue convergence theorem and Assumption~\eqref{eq:cond.small}.
  Here, the assumption $d \ge 3$ is essential.  In the $\eps$-independent
  case, Condition~\eqref{eq:cond.ew} is obsolete. The Main Lemma yields
  \begin{equation}
    \label{eq:upper.est.conf}
    \EWT k \Meps \le \EWN k X + \delta_k(\eps) 
  \end{equation}
  with $\delta_k(\eps) \to 0$ as $\eps \to 0$.
  
  To prove the opposite inequality we apply the Main Lemma once more, this
  time with $\HS_\eps := \Lsqr \Meps$, $q_\eps := q^\theta_{\Meps}$, $\HS' :=
  \Lsqr X$, $q' := q^\Neu_X$ and $\Phi_\eps u := u \restr X$ being the
  restriction operator.
  
  Again, Condition~\eqref{eq:cond.norm} is satisfied by
  Lemma~\ref{lem:no.conc.conf}, and the inequality in
  Condition~\eqref{eq:cond.quad} is trivially satisfied. Finally,
  Condition~\eqref{eq:cond.ew} is satisfied by the upper
  bound~\eqref{eq:upper.est.conf}. The Main Lemma therefore yields
  \begin{displaymath}
    \EWN k X \le \EWT k \Meps + \delta_k(\eps)
  \end{displaymath}
  with $\delta_k(\eps) \to 0$ as $\eps \to 0$. Together
  with Estimate~\eqref{eq:upper.est.conf} we are done.
\end{proof}

\subsection*{The Two-Dimensional Case}
In dimension $2$, the special form of the quotient
$q^\theta_\Meps(u)/\normsqr[\Meps] u$ causes a different behaviour. The
$\theta$-periodic eigenvalue of the Laplacian on a period cell $\Meps$ still
converges, but the limit depends on $\theta$, i.e., the $k$-th band
$B_k(\Mepsper)$ in general does not reduce to a point.  In particular, if we
assume that $\rho_\eps$ is monotonely decreasing (as $\eps \searrow 0$), the
first $\theta$-periodic eigenvalue is monotonely increasing (as $\eps \searrow
0$) since we have
\begin{equation}
  \label{eq:quot.dim2}
  \EWT k \Meps =  \inf_ u \frac{q^\theta_\Meps(u)}
                               {\normsqr[\Meps] u}
               =  \inf_ u \frac{\int_M |\dd u|^2}
                               {\int_M |u|^2 \rho_\eps^2s}
\end{equation}
due to the Min-max Principle and~\eqref{eq:conf.fact}. Here, the
infimum is taken over all $u \in \dom q^\theta_\Meps$ such that $u \ne 0$.
Note that $\dom q^\theta_\Meps$ is independent of $\eps$ as vector space.
Since the first band $B_1(\Mepsper)$ of a connected periodic manifold
$\Mepsper$ has always nontrivial interior (see Remark~\ref{rem:abs.cont}),
$B_1(\Mepsper)$ cannot reduce to a point as $\eps \to 0$.

As in the higher dimensional case, the norm on $\Meps$ converges to the norm
on $X$, i.e., the limit quadratic form lives in the Hilbert space $\Lsqr X$.
But there is no reason why the limit form should only be an integral over $X$
since the quadratic form $q^\theta_\Meps$ does not depend on $\eps$ any more
(see Equation~\eqref{eq:quot.dim2}).

Indeed, the following candidate for the limit form is the right one (a more
detailed motivation can be found in~\cite{post:00}, note that harmonic
functions minimize the energy integral, i.e., the integral over $|\dd u|^2$).
For $u \in \dom q^\Neu_X$ let $h = H^\theta u \in \dom q^\theta_M$ be the
$\theta$-periodic harmonic extension of $u$ on $M$. In particular, $u=h$ on
$X$ and $\laplacian {M \setminus X} h=0$ on $M \setminus X$ such that $h$ and
$\dd h$ are $\theta$-periodic, i.e., $h(x)=\theta(\gamma) h(\gamma x)$ resp.\ 
$\dd h(x)=\theta(\gamma) \dd h(\gamma x)$ for all $x \in M$ and $\gamma \in
\Gamma$ such that $\gamma x \in M$.  Then we set
\begin{displaymath}
  q^\theta_0(u):= \int_M | \dd (H^\theta u)|^2
\end{displaymath}
for all $u \in \dom q^\theta_0 := \dom q^\Neu_X$. Note that $q^\Neu_X(u) \le
q^\theta_0 (u)$ for all $u$. In particular, the corresponding operator to
$q^\theta_0$ has purely discrete spectrum denoted by $\EWT k 0$
(written in increasing order and repeated according to multiplicity).
Furthermore, since $\dom q^\theta_0 \supset \dom q^\Dir_X$ the Min-max
Principle yields
\begin{displaymath}
  \EWN k X \le \EWT k 0 \le \EWD k X.
\end{displaymath}

Now we show the convergence of the $\theta$-periodic eigenvalues on $\Meps$ to
the eigenvalues $\lambda^\theta_k(0)$:
\begin{theorem}
  \label{thm:conv.conf2}
  Suppose that $d=2$. Then $\EWT k \Meps \to \EWT k 0$ as $\eps \to 0$ for all
  $\theta \in \hat \Gamma$.
\end{theorem}

\begin{proof}
  First, we prove an upper bound. For this note that $\norm[X] u \le
  \norm[\Meps] {H^\theta u}$ and $q^\theta_0(u) = q^\theta_\Meps(H^\theta u)$.
  By the Min-max Principle (or formally one could also apply the Main Lemma)
  we obtain
  \begin{equation}
    \label{eq:upper.est.conf2}
    \EWT k \Meps \le \EWT k 0. 
  \end{equation}
  
  To prove the opposite inequality we apply the Main
  Lemma~\ref{lem:main.lemma} once more, this time with $\HS_\eps := \Lsqr
  \Meps$, $q_\eps := q^\theta_{\Meps}$, $\HS' := \Lsqr X$, $q' := q^\theta_0$
  and $\Phi_\eps u := u \restr X$ being the restriction operator.
  
  Again, Condition~\eqref{eq:cond.norm} is satisfied by
  Lemma~\ref{lem:no.conc.conf}. The inequality in
  Condition~\eqref{eq:cond.quad} is satisfied since 
  \begin{align*}
    q_\eps(u) &= 
    \int_X |\dd u|^2 + \int_{M \setminus X} |\dd u   |^2
    \\ &\ge
    \int_X |\dd u|^2 + 
             \int_{M \setminus X} |\dd (H^\theta (u \restr X))|^2  =
    q^\theta_0(\Phi_\eps u).
  \end{align*}
  Note that the harmonic function $H^\theta (u\restr X)$ minimizes the second
  integral, see Equation~\eqref{eq:harmonic}. Finally,
  Condition~\eqref{eq:cond.ew} is satisfied since $\EWT k \Meps \le \EWD k X$.
  The Main Lemma therefore yields
  \begin{displaymath}
    \EWT k 0 \le \EWT k \Meps + \delta_k(\eps)
  \end{displaymath}
  with $\delta_k(\eps) \to 0$ as $\eps \to 0$. With regard to
  Estimate~\eqref{eq:upper.est.conf2} the proof is finished.
\end{proof}

Next we characerise the domain of the operator $Q^\theta_0$ corresponding to
$q^\theta_0$. Note that $\dom Q^\theta_0$ consists of those $u \in \dom
q^\theta_0$ such that there exists a (unique) element $v \in L_2(X)$
satisfying
\begin{displaymath}
  q^\theta_0(u,w)= \iprod v w
\end{displaymath}
for all $w \in \dom q^\theta_0$. In particular, $Q^\theta_0 u = v$ (see
e.g.~\cite[Theorem~VI.2.1]{kato:66}). Here, we can give a more explicit
characterisation of the limit operator:
\begin{lemma}
  \label{lem:dom.op}
  The domain of the operator $Q_0^\theta$ corresponding to the limit quadratic
  form $q_0^\theta$ is given by
  \begin{equation}
    \label{eq:dom.op}
    \dom Q_0^\theta=\bigl\{ u \in \Sob[2] X \, \bigl| \bigr. \,\, 
         \normder u = \normder H^\theta u \text{ on $\bd X$} \bigr\}.
  \end{equation}
  Furthermore, $Q_0^\theta u = \Delta_X u$ for $u \in \dom Q_0^\theta$.
\end{lemma}
Here, $\normder u$ denotes the normal (outer) derivate with respect to $X$.
Furthermore, $\Sob[2] X$ denotes the Sobolev space of square integrable weak
derivatives up to second order.
\begin{proof}
  The lemma follows from the Gauss-Green Formula and the characterisation of
  $\dom Q^\theta_0$. Note that the integral over $\bd M$ vanishes since
  $h=H^\theta u$ and $\dd h$ are both $\theta$-periodic. Furthermore, $u,
  \Delta u \in L_2(X)$ imply $u \in \Sob[2] X$ by regularity theory.
\end{proof}
 
Since the limit operator is quite complicated, we are only able to construct
an example of a conformally deformed $2$-dimensional manifold with gaps in the
spectrum of its Laplacian:

\begin{example}
  \label{ex:dim.2}
  Let $\Mper :=\R \times \Sphere^1$ be a cylinder with $\Gamma=\Z$ acting on
  $\Mper$ by $\gamma (x,\sigma)=(\gamma+x,\sigma)$.  The periodic metric is
  given by $g=\dd x^2 + r^2 \dd \sigma^2$ for some fixed $r>0$. We choose
  $M=[0,1] \times \Sphere^1$ as period cell.
  
  Let $0<a<b<1$ and let $X=[a,b] \times \Sphere^1$ be the undisturbed region
  of $M$.  Note that normal coordinates exist on $\overline{M \setminus X}$.
  Let $\theta \in \hat \Gamma \cong \Torus^1$. In this context we prefer to
  view $\theta$ as $\e^{\im \theta} \in \Torus^1$.
  
  We first have to calculate the $\theta$-periodic harmonic extension
  $h=H^\theta u$ of a function $u \in \Ci {X}$ given by
  $u(x,\sigma)=v(x)\e^{\im n \sigma}$ for some $n \in \Z$, i.e., we have to
  solve the boundary value problem
  \begin{displaymath}
    \Delta_{M \setminus X} h =
     -\partial_{xx} h - \frac1{r^2} \partial_{\sigma \sigma} h= 0, 
     \qquad \text{with} \qquad
    \begin{aligned}
      h(a,\cdot) &= u(a,\cdot) \\ 
      h(b,\cdot) &= u(b,\cdot) \\ 
      h(1,\cdot) &= \e^{i\theta} h(0,\cdot)\\
      \partial_x h(1,\cdot) &= \e^{i\theta} \partial_x h(0,\cdot)
    \end{aligned}
  \end{displaymath}
  which has a unique solution. Next, we search for eigenvalues $\lambda=\EWT k
  0 \ge 0$ with eigenfunctions $u$. Again, by separating the variables we can
  calculate them expicitely. Since eigenfunctions $u$ have to be in the domain
  of $Q^\theta_0$, the normal derivatives of $u$ and the harmonic extension
  $h$ agree on $\bd X$ by the preceding lemma. Therefore we have to solve
  \begin{displaymath}
    \Delta_ X u =
     -\partial_{xx} u - \frac1{r^2} \partial_{\sigma \sigma} u = \lambda u, 
     \qquad \text{with} \qquad
    \begin{aligned}
      \partial_x u(a,\cdot) &= \partial_x h(a,\cdot) \\ 
      \partial_x u(b,\cdot) &= \partial_x h(b,\cdot).
    \end{aligned}
  \end{displaymath}
  This gives a restriction on the possible values of $\lambda$. A long, but
  straightforward calculation yields
  \begin{align}
    \label{eq:sol.0}
    2 \bigl( \cos (L \omega) - \cos \theta \bigr) &= 
      \ell \omega \sin (L \omega)\\
    \label{eq:sol.not.0}
    \Bigl( \omega^2 - \frac{n^2}{r^2} \Bigr) \sin (L \omega) &= 
    2 \omega \frac n r \cdot
    \frac{\cosh (\ell \frac n r) \cos (L \omega) - \cos \theta}
         {\sinh (\ell \frac n r)}
  \end{align}
  with $\lambda = \omega^2 +n^2/r^2$ and $\ell:=1-L=1-b+a$ where the first
  equation is valid for $n=0$ and $\omega \ge 0$ and the second equation for
  $n \ne 0$ and $\omega > 0$ (see~\cite{post:00}).  Note that $\ell$ is the
  length of the perturbed cylinder $M \setminus X$ and that $L$ is the length
  of the unperturbed cylinder $X$.
  
  If $n=0$ we obtain smooth functions $\theta \mapsto \omega_m(\theta)$ for
  each $m \in \N_0$ solving Equation~\eqref{eq:sol.0} (see
  Figure~\ref{fig:gaps3}). 
  \begin{figure}[h]
    \begin{center}
      \leavevmode
      \begin{picture}(0,0)(-7.5,0)
        \includegraphics{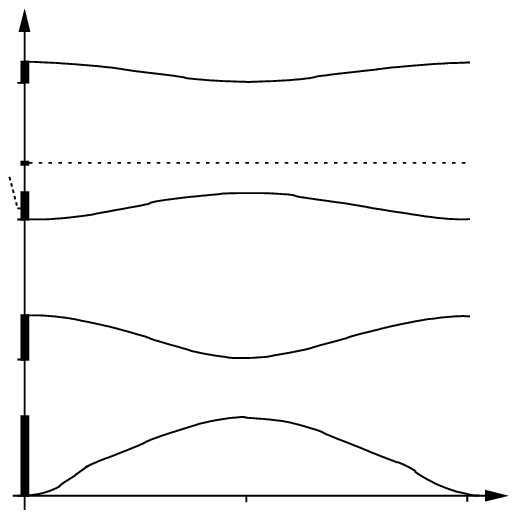}
      \end{picture}
      \setlength{\unitlength}{4144sp}
      \begin{picture}(2437,2453)(181,-1647)
        \put(213,-846){$\frac \pi L$}
        \put(181,-190){$\frac{2 \pi} L$}
        \put(2441,520){$\omega_4$}
        \put(2441, 67){$\eta_{1,1}$}
        \put(2441,-190){$\omega_3$}
        \put(2441,-648){$\omega_2$}
        \put(2441,-1352){$\omega_1$}
        \put(1343,-1605){$\pi$}
        \put(310,-1605){$0$}
        \put(2333,-1605){$2\pi$}
        \put(2618,-1482){$\theta$}
        \put(181,425){$\frac{3 \pi} L$}
        \put(407,401){$\sqrt{B_4}$}
        \put(407,165){$\sqrt{B_{1,1}}$}
        \put(407,-88){$\sqrt{B_3}$}
        \put(407,-788){$\sqrt{B_2}$}
        \put(407,-1234){$\sqrt{B_1}$}
        \put(407,665){$\sqrt \lambda$}
        \put(236,  1){$\frac 1 r$}
      \end{picture}
      \caption{The square root of the eigenvalues of the limit operator
          $Q_0^\theta$  plotted for $L=0.5$ and $r=1/13$. Here, at least $m=2$
          gaps occure.}
      \label{fig:gaps3}
    \end{center}
  \end{figure}
  In this case, we have $\lambda=\sqrt{\omega_m(\theta)}$.  Furthermore, note
  that the compact intervals $B_m:=\{\omega^2_m(\theta); \, 0 \le \theta \le
  2\pi \}$ are all disjoint: One can prove that
  \begin{align*}
    \inf B_m &= \Bigl(\frac{m\pi} L \Bigr)^2 & \text{and} &&
    \sup B_m &= \Bigl(\frac{(m+1)\pi} L \Bigr)^2 - \eps_0
  \end{align*}
  for all $m \in \N_0$ if $\eps_0=\eps_0(L)$ is small enough.  Finally
  note that $(m\pi/L)^2$ are the Neumann eigenvalues of the interval $[0,L]$.
  
  If $n \ne 0$ we have 
  \begin{displaymath}
    \eta := 
    \sqrt {\lambda} = 
    \sqrt{\omega^2 + n^2/r^2} >
    \frac {|n|} r \ge 
    \frac 1 r.
  \end{displaymath}
  If we replace $\omega$ by $\sqrt{\eta^2 - n^2/r^2}$ in
  Equation~\eqref{eq:sol.not.0} we obtain solutions $\theta \mapsto
  \eta_{n,p}(\theta)$ for $n,p \in \N$ (see Figure~\ref{fig:gaps3}).  Note
  that we do not expect that the intervals $B_{n,p}:=\{\eta^2_{n,p}(\theta) ;
  \, 0 \le \theta \le 2\pi \}$ are disjoint, we rather expect that the
  intervals $B_{n,p}$ cover the gaps between the intervals $B_m$ when $m,n$ or
  $p$ are large. But we still have $\inf B_{n,p} \ge 1/r^2$, i.e., if $r \le
  \frac L {m \pi}$, the intervals $B_0, \dots, B_m$ remain disjoint.
  Therefore we have proven the following:
  \begin{theorem}
    \label{thm:dim.2}
    Suppose that $\Mepsper$ is a conformally perturbed cylinder of radius
    $r>0$ with conformal factor satisfying Conditions~\eqref{eq:cond.one}
    and~\eqref{eq:cond.small}. Suppose further that the unperturbed area is a
    cylinder $X$ of length $0 < L < 1$ (periodically continued with period
    $1$). Then the corresponding Laplacian has at least $m$ gaps if $r \le
    \frac L {m \pi}$ and if $\eps>0$ is small enough.
  \end{theorem}
\end{example}

\section{Conclusions and outlook}
So far we have proven the existence of two different classes of periodic
manifolds with spectral gaps. But we still have not found a satisfying answer
(in terms of geometrical properties) whether a given periodic manifold has
gaps or not. We define the $\nu$-isoperimetric constant of a periodic manifold
$\Mper$ as
\begin{displaymath}
  I_\nu(\Mper) := 
     \inf_M \inf_\Omega \frac{(\vol_{d-1} \bd \Omega)^\nu}
                             {(\vol_d \Omega)^{\nu-1}}
\end{displaymath}
where $\vol_k$ denotes $k$-dimensional Riemannian volume and $\nu>1$. The
infimum is taken over all period cells $M$ of $\Mper$ and all open
submanifolds $\Omega \subset M$ with $\bd M \cap \bd \Omega = \emptyset$.  In
both classes of examples given in Sections~\ref{sec:constr} and~\ref{sec:conf}
we have $\vol_{d-1}(\bd \Meps) \to 0$ whereas $\vol_d(\Meps)$ is bounded from
below by some positive constant. Therefore $I_\nu(\Mepsper) \to 0$ as $\eps
\to 0$ for all $\nu>1$.  Are the isoperimetric constants some hint for the
analytic decoupling? Here, by analytic decoupling we mean that the $k$-th
$\theta$-periodic eigenvalue of $\Meps$ converges uniformly to some
$\theta$-independent constant $\lambda_k$ (as in
Theorems~\ref{thm:conv.constr},~\ref{thm:conv.cyl} and~\ref{thm:conv.conf}).
Note, for example that the isoperimetric constants do not all converge to $0$
for a straight cylinder $\Sphere^{d-1} \times \R$ of diameter of order $\eps$
and length of a period cell of order $\eps^\alpha$, $\alpha>0$. Clearly, the
spectrum of the straight cylinder has no gaps.

What r\^ole does the curvature play? In the construction in
Section~\ref{sec:constr} as well as the conformal deformation in
Section~\ref{sec:conf} the curvatures are neither bounded from below nor
bounded from above. There is a result of Li \cite{li:94} showing that the
essential spectrum of a complete non-compact Riemannian manifold $\Mper$ with
non-negative Ricci curvature and a pole is equal to $[0,\infty)$. (A pole is a
point $x_0$ where the exponential map is a diffeomorphism from $T_{x_0}\Mper$
onto $\Mper$, which is a strong geometric condition.) If furthermore, $\Mper$
is periodic then the bound on the curvature prevents the existence of spectral
gaps. In contrast, we would expect to have a great number of gaps if the
curvature has great absolute values.


\section*{Acknowledgment}
  I am indebted to my thesis advisor Rainer Hempel for his kind attention.
  Furthermore I thank Colette Ann\'e for the helpful discussion concerning her
  article. After all I thank Fernando Lled\'o for final comments.


\providecommand{\bysame}{\leavevmode\hbox to3em{\hrulefill}\thinspace}




\begin{thebibliography}{10}

\bibitem{anne:87}
C.~Ann\'e, \emph{{Spectre du Laplacien et \'ecrasement d'anses}}, Ann. Sci.
  \'Ec. Norm. Super., IV. S\'er. \textbf{20} (1987), 271--280.

\bibitem{atiyah:76}
M.~F. Atiyah, \emph{{Elliptic operators, discrete groups and von Neumann
  algebras}}, Asterisque \textbf{32-33} (1976), 43--72.

\bibitem{bratteli:99}
O.~Bratteli, P.~E.~T. J{\o}rgensen, and D.~W. Robinson, \emph{Spectral
  asymptotics of periodic elliptic operators}, Math. Z. \textbf{232} (1999),
  no.~4, 621--650.

\bibitem{bruening:92}
J.~Br{\"u}ning and T.~Sunada, \emph{{On the spectrum of periodic elliptic
  operators}}, Nagoya Math. J. \textbf{126} (1992), 159--171.

\bibitem{chavel:93}
I.~Chavel, \emph{{Riemannian geometry}}, Cambridge University Press, Cambridge,
  1993.

\bibitem{chavel-feldman:78}
I.~Chavel and E.~A. Feldman, \emph{{Spectra of domains in compact manifolds}},
  J. Funct. Anal. \textbf{30} (1978), 198--222.

\bibitem{chavel-feldman:81}
I.~Chavel and E.~A. Feldman, \emph{{Spectra of manifolds with small handles}},
  Comment. Math. Helv. \textbf{56} (1981), 83--102.

\bibitem{davies:96}
E.~B. Davies, \emph{{Spectral theory and differential operators}}, Cambridge
  University Press, Cambridge, 1996.

\bibitem{davies-harrell:87}
E.~B. Davies and E.~M.~II Harrell, \emph{{Conformally flat Riemannian metrics,
  Schr{\"o}dinger operators, and semiclassical approximation}}, J. Diff.
  Equations \textbf{66} (1987), 165--188.

\bibitem{donnelly:81}
H.~Donnelly, \emph{{On $L\sp 2-$Betti numbers for Abelian groups}}, Can. Math.
  Bull. \textbf{24} (1981), 91--95.

\bibitem{exner-seba:89}
P.~Exner and P.~{\v S}eba, \emph{{Bound states in curved quantum waveguides}},
  J. Math. Phys. \textbf{30} (1989), no.~11, 2574--2580.

\bibitem{froese-herbst:00}
R.~Froese and I.~Herbst, \emph{{Realizing holonomic constraints in classical
  and quantum mechanics}}, Studies in Advanced Mathematics \textbf{16} (2000),
  121--131.

\bibitem{fukaya:87}
K.~Fukaya, \emph{{Collapsing of Riemannian manifolds and eigenvalues of Laplace
  operator}}, Invent. Math. \textbf{87} (1987), 517--547.

\bibitem{green:97}
E.~L. Green, \emph{{Spectral theory of Laplace-Beltrami operators with periodic
  metrics}}, J. Diff. Equations \textbf{133} (1997), no.~1, 15--29.

\bibitem{hempel:92b}
R.~Hempel, \emph{{Second order perturbations of divergence type operators with
  a spectral gap}}, Operator Theory: Advances and Applications \textbf{57}
  (1992), 117--126.

\bibitem{hempel-herbst:95a}
R.~Hempel and I.~Herbst, \emph{{Bands and gaps for periodic magnetic
  Hamiltonians}}, Operator Theory: Advances and Applications \textbf{78}
  (1995), 175--184.

\bibitem{hempel-herbst:95b}
\bysame, \emph{{Strong magnetic fields, Dirichlet boundaries, and spectral
  gaps}}, Commun. Math. Phys. \textbf{169} (1995), no.~2, 237--259.

\bibitem{hempel-lienau:00}
R.~Hempel and K.~Lienau, \emph{{Spectral properties of periodic media in the
  large coupling limit}}, Commun. Partial Differ. Equations \textbf{25} (2000),
  1445--1470.

\bibitem{kato:66}
T.~Kato, \emph{{Perturbation theory for linear operators}}, Springer-Verlag,
  Berlin, 1966.

\bibitem{sunada:89}
T.~Kobayashi, K.~Ono, and T.~Sunada, \emph{{Periodic Schr{\"o}dinger operators
  on a manifold}}, Forum Math. \textbf{1} (1989), no.~1, 69--79.

\bibitem{li:94}
J.~Li, \emph{{Spectrum of the Laplacian on a complete Riemannian manifold with
  nonnegative Ricci curvature which possess a pole}}, J. Math. Soc. Japan
  \textbf{46} (1994), no.~2, 213--216.

\bibitem{lott:01}
J.~Lott, \emph{On the spectrum of a finite-volume negatively-curved manifold},
  Amer. J. of Math. \textbf{123} (2001), 185--205.

\bibitem{mitchell:01}
K.~A Mitchell, \emph{{Gauge fields and extrapotentials in constrained quantum
  systems}}, Phys. Rev. A \textbf{63} (2001), Nr.~042112.

\bibitem{post:00}
O.~Post, \emph{Periodic manifolds, spectral gaps, and eigenvalues in gaps},
  Ph.D. thesis, Technische Universit{\"a}t Braunschweig, 2000.

\bibitem{reed-simon-4}
M.~Reed and B.~Simon, \emph{{Methods of modern mathematical physics IV:
  Analysis of operators}}, Academic Press, New York, 1978.

\bibitem{reed-simon-1}
\bysame, \emph{{Methods of modern mathematical physics I: Functional
  analysis}}, Academic Press, New York, 1980.

\bibitem{skriganov:79}
M.~M. Skriganov, \emph{{Proof of the Bethe-Sommerfeld conjecture in dimension
  two}}, Sov. Math., Dokl. \textbf{20} (1979), 956--959.

\bibitem{yoshitomi:98}
K.~Yoshitomi, \emph{{Band gap of the spectrum in periodically curved quantum
  wave\-guides}}, J. Diff. Equations \textbf{142} (1998), no.~1, 123--166.

\end{thebibliography}
\end{document}